\documentstyle[12pt,epsf,twoside,fleqn,espcrc1]{article}
\title{Flow systematics from SIS to SPS energies}
\author{Jean-Yves Ollitrault\address{Service de physique th\'eorique, 
C.E. Saclay,F-91191 Gif-sur-Yvette cedex}
\thanks{Member of CNRS.}}
\begin{document}
\maketitle
\begin{abstract}
The various flow phenomena observed at energies between 50~AMeV and 160~AGeV 
are reviewed. 
I first define three types of flow: directed flow and elliptic flow, 
which are the two first Fourier components of the azimuthal 
distribution in non-central collisions; 
radial flow, which is deduced from an analysis of transverse momentum 
spectra in central collisions. 
Then, I review the observations of directed flow and elliptic flow, with 
emphasis on recent results. I discuss their dependence on various parameters:
global geometry (impact parameter, mass numbers of colliding 
nuclei and bombarding energy) and individual observables 
(rapidity, transverse momentum and particle type). 
Finally, I explain how azimuthal distributions can be measured experimentally. 
\end{abstract}

\section{THREE TYPES OF FLOW}

The azimuthal angles of particles emitted in a nuclear collision are 
correlated with the direction of impact parameter \cite{Gust84}. 
In this talk, I use the generic term ``flow'' to refer to these correlations 
(which are essentially macroscopic effects),
irrespective of any theoretical interpretation. 

The identification of the direction of impact parameter in a high energy 
collision is a non trivial task, which relies on a study of azimuthal 
correlations between the reaction products (see Section 4). 
In practice, this identification is possible only if the system is large 
enough: in proton-proton and proton-deuteron collisions at 28~GeV, 
the observed azimuthal correlations are back to back and can be explained
by momentum conservation alone \cite{Fost72}. 
In proton-nucleus collisions, flow may be present, as indicated by a 
recent study of correlations between pions and nucleons \cite{Awes96}. 
However, flow is seen unambiguously only in nucleus--nucleus collisions, 
where it was discovered at Bevalac in 1984 \cite{Gust84}. 
It is thought to be highly sensitive to medium properties, 
and has therefore been extensively studied in recent years.

\begin{figure}[htb]
\centerline{\epsfbox{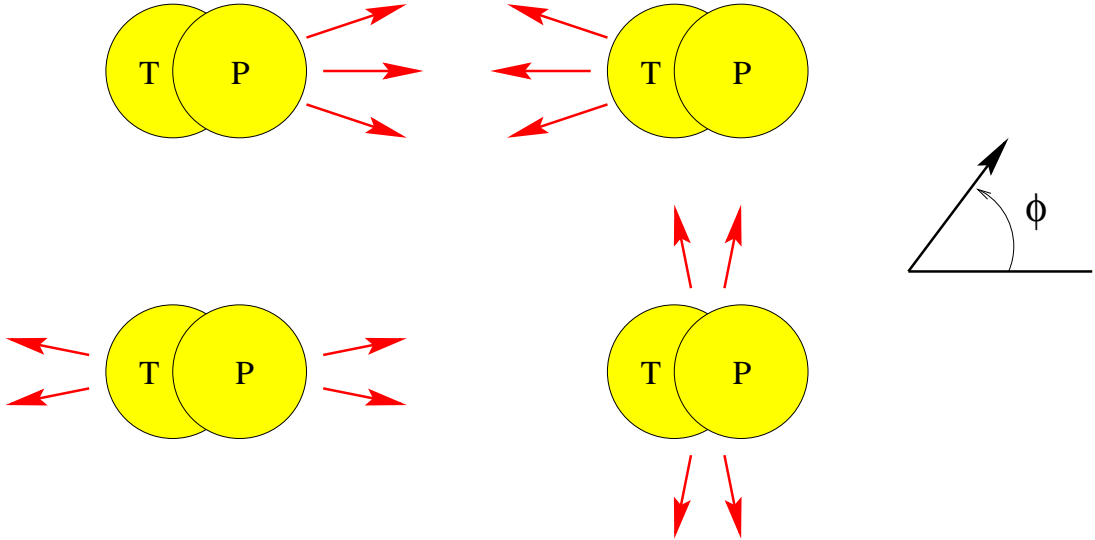}}
\caption{The four major types of azimuthal anisotropies, viewed in the 
transverse plane. The target is denoted by T and the projectile by P. 
Top: Directed flow in the projectile rapidity region, positive (left) 
and negative (right). In the target rapidity region, the left and right 
figures are  interchanged. Bottom: elliptic flow, in-plane (left) and 
out-of-plane (right).}
\label{fig:fig1}
\end{figure}

In this section, I denote by $\phi$ the azimuthal angle of an outgoing 
particle, measured from 
the direction of impact parameter (see Fig.\ref{fig:fig1}). 
There is flow if the $\phi$ distribution is not isotropic. 
A convenient measure of flow is provided by the Fourier coefficients, 
$\langle\cos n\phi\rangle$ (where the brackets  denote 
an average value in a given kinematic $(p_T,y)$ window). 
The two major flow effects observed so far correspond to the 
$n=1$ \cite{Gust84} and $n=2$ \cite{Gutb89} Fourier components. 
These effects, which I shall name {\sl directed\/} and 
{\sl elliptic\/} flow respectively, are illustrated in Fig.\ref{fig:fig1}. 
In the projectile rapidity region, directed flow is 
{\sl positive\/} if $\langle\cos\phi\rangle >0$, and 
{\sl negative\/} if $\langle\cos\phi\rangle <0$. 
For symmetric collisions, $\langle\cos\phi\rangle$ is an odd function of 
the centre-of-mass rapidity, and signs are therefore 
reversed in the target rapidity region. 
Similary, for elliptic flow, I distinguish 
{\sl in-plane\/} elliptic flow if $\langle\cos 2\phi\rangle >0$, and 
{\sl out-of-plane\/} elliptic flow if $\langle\cos 2\phi\rangle <0$. 
Note that unlike directed flow, elliptic flow has the same sign 
in the target and projectile rapidity regions, at least for symmetric 
systems. 

In central collisions, the direction of impact parameter cannot be defined: 
azimuthal distributions become flat, and directed and elliptic flow vanish. 
However, 
a third type of flow, called {\sl radial flow\/}, has been defined for central 
collisions (although it is not a ``flow'' in the sense defined above)
\cite{Jeon94,Lisa95,Brau95}. 
The idea is to compare the measured kinetic energy (or transverse mass in 
relativistic collisions) spectra to those of a thermal fluid expanding  
outwards \cite{Siem79,Schn93}. 
Consider, for simplicity, a non-relativistic perfect fluid with temperature 
$T$ and velocity ${\bf v}_{\rm flow}$. The velocity of a particle in the fluid 
can be written as ${\bf v}={\bf v}_{\rm flow}+{\bf v}_{\rm th}$, 
where ${\bf v}_{\rm th}$ is the velocity of the random thermal motion. 
Then the average kinetic energy of a particle with mass $m$ is 
$\langle E\rangle =mv_{\rm flow}^2/2 + 3T/2$: while the thermal energy 
is independent of $m$, the collective flow contribution is proportional 
to $m$. From the spectra of particles with different masses, 
one can thus extract both the thermal and the collective component. 
This will not be discussed further here. 

\begin{figure}[htb]
\begin{minipage}[t]{81mm}
\epsfbox{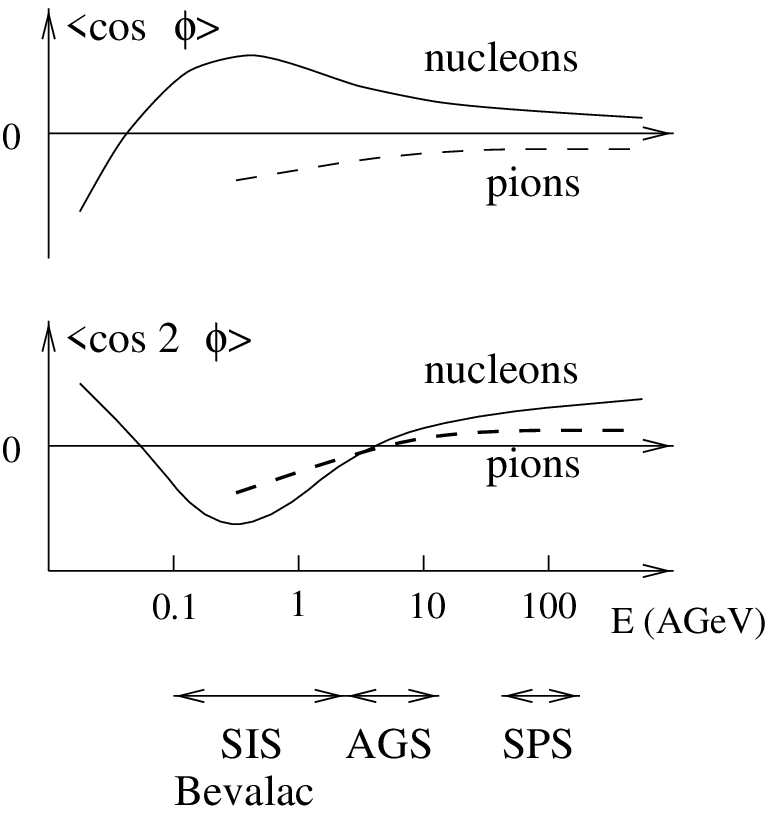}
\caption{Schematic behaviour of the magnitudes of directed flow 
(top) and elliptic flow (bottom) as a function of the bombarding kinetic 
energy per nucleon in the laboratory frame. Full lines: proton flow; 
dashed lines: pion flow.}
\label{fig:fig2}
\end{minipage}
\hspace{\fill}
\begin{minipage}[t]{73mm}
\epsfbox{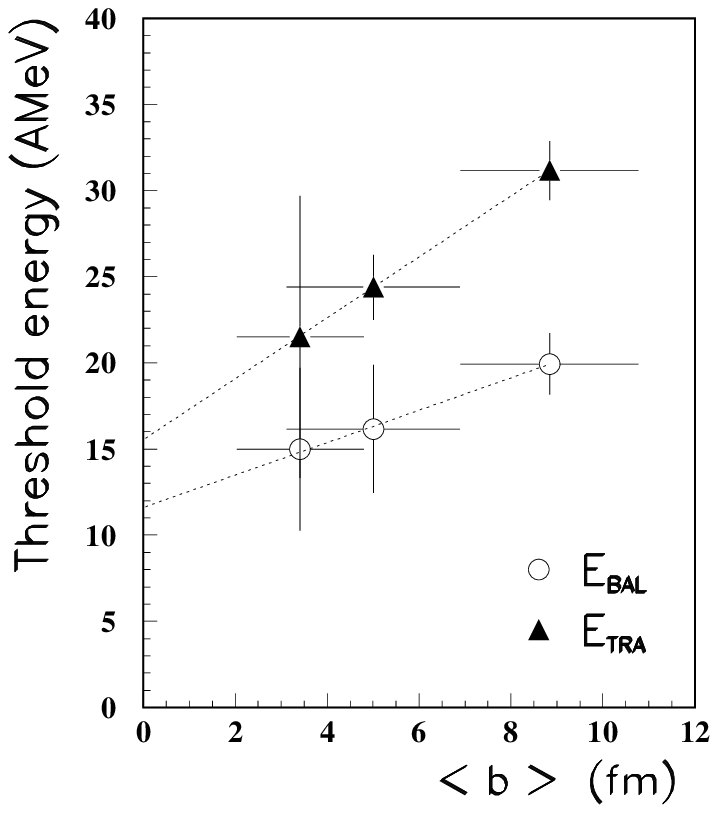}
\caption{Energies of vanishing directed flow ($E_{BAL}$) and elliptic 
flow ($E_{TRA}$), as a function of impact parameter 
(from [15]), in the centre of mass system. 
The corresponding laboratory energies are larger by a factor of 4.}
\label{fig:fig3}
\end{minipage}
\end{figure}

Fig.\ref{fig:fig2} displays the typical variation of directed and 
elliptic flow with the bombarding energy. The high energy 
part will be discussed in detail in the next two sections. 
Here, I comment on the change of sign of both directed and elliptic 
flow below 100~AMeV. 
At low energies, the interaction is dominated by the attractive nuclear
mean field, which as two effects: first, projectile nucleons are 
deflected towards the target, resulting in negative directed flow 
\cite{Tsan86}; second, the projectile and target form a rotating system, 
and the centrifugal force emits particles in the rotation plane 
(``rotation-like effect'' \cite{Wils90}), producing in-plane elliptic 
flow \cite{Chit86}. 
At higher energies, individual nucleon-nucleon collisions dominate 
over mean field effects. They produce a positive pressure, 
which deflects the projectile and intermediate rapidity fragments 
away from the target (``bounce-off'' and ``sidesplash'' effects \cite{Gust84}),
resulting in positive directed flow. Furthermore, the participant nucleons,  
which are compressed in the region where the target and the projectile overlap 
(see Fig.\ref{fig:fig1}), cannot escape in the reaction plane due to the 
presence of the spectator nucleons (``squeeze-out effect'' \cite{Gutb89}), 
producing out-of-plane elliptic flow. 

The crossing energies at which directed \cite{Krof89} and elliptic 
\cite{Pope94} flow cancel are displayed in Fig.3 for Au--Au collisions,
as a function of impact parameter. An extrapolation to zero impact parameter 
gives values which are consistent, within error bars, with the threshold 
energy for radial flow, estimated as $E_{\rm RAD}=8.7\pm 2.5$~AMeV. 
As Crochet et al. point out \cite{Croc97a}, this suggests that a common 
mechanism is at the origin of positive directed flow, out-of-plane elliptic 
flow and radial flow: the three phenomena appear when the attractive 
forces are counterbalanced by the thermal pressure. 
A possible relation of this crossing phenomenon 
with the nuclear liquid--gas phase transition 
is discussed in \cite{Bona87}. 

I now turn to a more detailed analysis of directed and elliptic flow 
at energies above 100~MeV per nucleon. 

\section{DIRECTED FLOW}

The positive directed flow of nucleons has been extensively studied below 
2~GeV per nucleon, both theoretically and experimentally. 
Since it is essentially a pressure effect, 
it is naturally expected to be sensitive to the compressibility of 
nuclear matter. The momentum dependence of nuclear interactions is also 
important \cite{Zhan94}. 

The magnitude of directed flow depends on the centrality of the 
collision: it vanishes for central collisions by symmetry, 
and for very peripheral collisions where compressional effects are 
low. It is therefore maximum for semicentral collisions \cite{Doss86}, 
at an impact parameter of the order of 5--6~fm for Au--Au collisions
\cite{Croc97b}. The measurements of directed flow to be discussed below 
are made at the centrality where it is largest. 

As mentioned in Section 1, directed flow is an odd function of the 
centre-of-mass rapidity in symmetric collisions.
It is therefore linear near mid-rapidity. Furthermore, a saturation
is observed near projectile and target rapidities, resulting in a typical 
S shape \cite{Dani85}. 

Due to this peculiar dependence, and to acceptance considerations, 
the strength of directed flow is usually measured by plotting the 
average momentum projected on the direction of impact parameter, 
$\langle p_T\cos\phi\rangle$, as a function of rapidity, and by 
taking the slope of this curve near mid-rapidity, the so-called flow 
parameter \cite{Doss86}: 
\begin{equation}
\label{F}
F = y_{\rm proj} \left.{d\langle p_T\cos\phi\rangle\over dy}
\right|_{y=y_{\rm cm}}
\end{equation}
where $y_{\rm proj}$ denotes the projectile rapidity. 

The flow parameter of protons scales with the mass numbers $A$ and $B$ of the 
colliding nuclei approximately like $A^{1/3}+B^{1/3}$ \cite{Chan97}.  
This is an important point. 
If the system behaved like a perfect fluid, it would obey geometric scaling 
\cite{Bala84} and the value of $F$ would be independent 
of $A$ for a symmetric collision (neglecting the nuclear skin depth). 
The fact that $F$ is proportional to $A^{1/3}+B^{1/3}$, i.e. to the 
collision time, which was predicted by microscopic calculations \cite{Lang91}, 
is an indication that the system is on its way towards thermalisation, 
but only partially thermalised \cite{Dani88}. 

Further insight into the mechanism of directed flow is obtained by 
studying the production of fragments.
Their directed flow increases with their mass at 
Bevalac and SIS energies \cite{Croc97b,Doss87}. 
A similar behaviour at AGS energies was reported at this 
conference \cite{Volo97b}. 
Repeating the simple fluid dynamical argument given in Section 1 for 
radial flow, this suggests that directed flow is an effect of 
collective motion in the expanding system. 

The variation of directed flow with bombarding energy is also instructive: 
between 150~MeV and 1~GeV per nucleon, i.e. in the non relativistic regime 
(recall that energies in the centre of mass are 4 times smaller), 
the flow parameter $F$  is approximately proportional to the projectile 
momentum \cite{Herr97}. In other terms, the velocities of outgoing particles
scale like the incoming velocity $v_i$. Although this scaling was predicted 
in a hydrodynamical model \cite{Bala84}, it is not an effect of 
thermalisation. It simply reflects the fact that in this energy range, the 
only relevant velocity scale is $v_i$: 	effects associated with 
the Fermi velocity $v_F$ or with relativity are negligible 
in the energy regime where $v_F\ll v_i\ll c$.

At lower energies, attractive effects become important, and  
the scaling breaks down. Furthermore, since the attractive part is essentially
a surface effect, the $A$ dependence becomes more complicated 
\cite{Soff95}.

At relativistic energies, the behaviour is also different. 
The scaled flow $F_s\equiv F/(A^{1/3}+B^{1/3})$, whose value below 1~AGeV 
is approximately $F_s\simeq 0.03\, p_{\rm proj}$, 
reaches a maximum of 50~MeV and then decreases slowly. 
A first indication of this decrease between 1 and 2~AGeV was reported 
at the last Quark Matter conference \cite{Herr97}. It is confirmed 
between 2 and 6 AGeV by results presented at this conference  \cite{Liu97}. 
At 10.8~AGeV, the E877 collaboration measures $F_s\simeq 35$~MeV 
\cite{Barr97a}, compatible with the value $F_s\simeq 31\pm 7\pm 6$~MeV
found by E917 \cite{Ogil97}. 

The first evidence of directed flow at SPS was reported by the WA98 
collaboration at the Jaipur Conference this year \cite{Nish97}. 
More recent measurements have been made by the NA49 \cite{Appe97} and 
NA45 \cite{Cere97} collaborations. 
Directed flow seems to be concentrated in the fragmentation regions, 
and almost absent in the central rapidity region: 
the rapidity dependence does not follow an S-shaped curve 
(deviations from this shape are already seen at AGS by E917 \cite{Ogil97}) 
and the flow parameter, defined as the slope near mid-rapidity 
(Eq.(\ref{F})), becomes irrelevant. 
The value of $\langle p_T\cos\phi\rangle$ in the fragmentation region 
is at least three times smaller than at AGS \cite{Nish97}, 
confirming the decrease discussed above. 

The flow parameter $F$ is a $p_T$--integrated observable. 
A more detailed information on directed flow can be obtained by plotting 
$\langle\cos\phi\rangle$ as a function of $p_T$.
At low $p_T$, $\langle\cos\phi\rangle$ varies linearly with $p_T$. 
This is because the momentum distribution must be a regular 
function of $p_x=p_T\cos\phi$ and $p_y=p_T\sin\phi$.
It was recently pointed out \cite{Volo97a,Volo97b} that directed flow can be 
{\sl negative\/} at low $p_T$ if strong radial flow occurs in the 
system. 
At higher $p_T$, the variation of directed flow gives an information 
on momentum dependent interactions \cite{Pan93,Li96a}. 

Directed flow of particles other than nucleons has been observed, but 
the available information is less detailed. 
Pions exhibit weak negative directed flow (i.e. opposite to the nucleons), 
first observed in asymmetric collisions \cite{Goss89,Adya90}. 
This is interpreted as a consequence of pion rescattering 
on spectator matter ($\pi N\to\pi N$) \cite{Bass93a}. 
The same effect has been recently observed at AGS \cite{Barr97b}
(although the flow is positive for high $p_T$, high rapidity pions)
and at SPS \cite{Awes96,Nish97,Appe97}. 

Directed flow of strange particles has been also studied at Bevalac and 
SIS. While $\Lambda$ baryons exhibit positive directed flow, of 
roughly the same magnitude as protons \cite{Just95},  
kaon directed flow is compatible with zero \cite{Ritm95}. 
This may seem surprising since only associate production is kinematically 
allowed at these energies (2~AGeV), and $\Lambda$'s and $K$'s 
are therefore produced in pairs. 
Since $K^+$'s have a large mean free path in nuclear matter 
(unlike pions, they 
cannot be absorbed), their flow provides a clean probe for the kaon potential
in a nuclear medium \cite{Li95}. 
First data on $K^+$ directed flow at AGS were presented at this conference
\cite{Volo97b}. The value is also consistent with 0, except for low-$p_T$ 
kaons where it is negative, i.e. opposite to the proton flow, 
which can be interpreted as an effect of Coulomb repulsion. 

The first observation of antiproton directed flow was reported at 
this conference \cite{Kaba97}, but the sign is unclear yet.  
Negative flow is expected, due to $p\bar p$ annihilation \cite{Jahn94}. 

Finally, the isospin dependence of directed flow may give interesting 
results: 
Coulomb effects, among others \cite{Li96b}, are expected to produce 
a small isospin dependence which, if measured, could provide a direct 
information on the size of the system. 
Experimentally, this can be studied in two different ways: 
first, one can compare the magnitude of proton directed flow in 
colliding systems with different isospins. This has been done recently 
at low energies\cite{Pak97}. 
Alternatively, one can compare the magnitude of directed flow for 
particles of a given isospin multiplet. 
No obvious difference is seen between neutron and proton 
\cite{Elaa94}, or between $K^+$ and $K^0_s$  \cite{Ritm95}. 
A difference between $\pi^+$ and $\pi^-$ flows at very 
low transverse momenta is seen by E877 \cite{Barr97b}. 

\section{ELLIPTIC FLOW}

While directed flow keeps the same sign from SIS to SPS energies, 
a transition from out-of-plane to in-plane elliptic flow occurs at 
relativistic energies, which was predicted a few years ago 
\cite{Olli92,Olli93a,Olli93b} and recently observed \cite{Barr97a}. 
In this section, 
I first recall a few general properties of elliptic flow. Then, 
I review the observations at SIS energies, and the recent 
measurements at AGS and SPS. 

The impact parameter dependence of elliptic flow is similar to that of 
directed flow \cite{Tsan96}. 
However, the maximum is found at a slightly higher impact parameter, 
about 7~fm for Au--Au collisions \cite{Bast97}, instead of 5-6~fm for 
directed flow. 
On the other hand, contrary to directed flow, the rapidity dependence 
is essentially flat at SIS and AGS \cite{Gutb90,Barr97a}. At SPS, 
a maximum is observed at mid-rapidity for protons \cite{Appe97}. 

Elliptic flow is higher for fragments than for protons 
\cite{Bast97,Gutb90,Bril96}: as for directed flow, this is a signature 
of collective motion. 
However, the magnitude of elliptic flow is larger for larger nuclei 
\cite{Gutb90,Alar96}, which indicates that only partial thermalisation 
is achieved. 

Elliptic flow is always out-of-plane at Bevalac and SIS energies. 
It reaches a maximum at about 400~AMeV \cite{Gutb90,Bril96,Lamb94}. 
However, the variation between 250 and 800 AMeV is very weak \cite{Bast97}, 
which suggests the same scaling behaviour as observed for directed flow 
in this energy region. 
This scaling can also be seen by studying the $p_T$ dependence of 
elliptic flow. At low $p_T$, simple mathematical arguments (similar 
to those used for directed flow) show that it is proportional to $p_T^2$ 
\cite{Dani95}. Then it rises as a function of $p_T$, and a universal 
curve is obtained when $p_T$ is scaled by the projectile momentum 
$p_{\rm proj}$ \cite{Bast97,Lamb94}. 

Charged \cite{Bril93} and neutral \cite{Vene93} pions also undergo 
out-of-plane elliptic flow. The dominant contribution to this effect 
is absorption on nucleons ($\pi NN\to\Delta N\to NN$) \cite{Bass93b}. 
Concerning the isospin dependence, 
the magnitudes are similar for neutrons and protons \cite{Lamb94,Leif93}, 
and no difference is seen between $\pi^+$ and $\pi^-$ \cite{Bril97}. 

The transition from out-of-plane to in-plane elliptic flow at 
relativistic energies is a consequence of the Lorentz contraction of 
the spectators. They leave the interaction region after a time of 
the order $2R/\gamma$ ($R$ being the nuclear radius and $\gamma$ the 
Lorentz contraction factor). Later, outgoing particles are 
free to escape anywhere in the transverse plane. However, if partial 
thermalisation occurs, pressure will drive them along the 
lines of pressure gradient (recall that the force per unit volume is 
$-{\bf\nabla}P$). Since the overlap region between target and 
projectile has a smaller size along the impact parameter (see Fig.1), 
the pressure gradient is larger in the reaction plane \cite{Olli92}. 
This produces in-plane elliptic flow. 

First indications of non-zero elliptic flow were obtained a few years ago 
at AGS by E877 \cite{Barr94} and at SPS by WA93 \cite{Viyo95} and 
NA49 \cite{Wien96}. However, conclusive evidence that it was 
in-plane came only recently \cite{Barr97a}. 
At AGS, in-plane elliptic flow is seen 
for protons, but not for pions \cite{Barr97b}, while it is seen for 
both pions and protons at SPS \cite{Nish97,Appe97}. 

From the qualitative arguments given above, one expects that the 
transition to in-plane elliptic flow occurs when the Lorentz 
contraction becomes significant \cite{Olli93a}. 
The first measurement of the transition energy was reported at this 
conference by the E895 collaboration, who obtain a value lying 
between  4 and 6~AGeV \cite{Liu97}. 
At these energies, the kinetic energy is almost equal to the mass 
energy in the centre of mass system, 
thus confirming the validity the above argument. 
A more thorough analysis has recently shown that the transition energy 
reflects the time dependence of the pressure \cite{Sorg97a}. 

Let us finally comment on the magnitude of in-plane elliptic flow: 
the integrated value for protons increases approximately from 2\%  to 6\%  
between AGS \cite{Barr97a} and SPS \cite{Appe97}. 
RQMD calculations \cite{Sorg97b} predict values of the 
order of 4\%  at SPS, in qualitative agreement with experimental data
\cite{Appe97,Cere97}, but significantly lower than the predictions of a 
hydrodynamical calculation \cite{Olli92}. This suggests that 
that the colliding system is not fully thermalised locally. 
The variation of in-plane elliptic flow with the number of collisions 
per particle, i.e. with the degree of thermalisation achieved in 
the system, is studied in \cite{Fili97}. 

\section{FLOW ANALYSIS}

I now briefly explain how the results presented in the previous
sections can be obtained experimentally. 
First, one must estimate the direction of impact parameter event by event. 
If positive directed flow is present, adding the transverse momenta
of all particles detected in the projectile rapidity region yields 
a vector which points in the direction of impact parameter 
(see Fig.\ref{fig:fig1}). This forms the basis of 
the transverse momentum method \cite{Dani85}. 
More generally, the azimuthal angle $\Phi_R$ of the impact parameter 
(measured from a fixed point in the detector) 
is estimated by constructing the following vector in the transverse plane:
\begin{equation}
\label{Q}
{\bf Q} =\pmatrix{Q\,\cos\Phi_R\cr Q\,\sin\Phi_R}=
\sum_{k=1}^N {w_k\pmatrix{\cos\phi_k\cr\sin\phi_k}}
\end{equation}
where the sum runs over detected particles with azimuthal angles 
$\phi_k$, and $w_k$ is a weight 
which is positive (resp. negative) in the projectile (resp. target)
rapidity region. 
$w_k$ can be any function of $p_T$ and $y$, and must be optimised to 
yield the best possible reaction plane determination. 

\begin{figure}[htb]
\begin{minipage}[t]{82mm}
\epsfbox{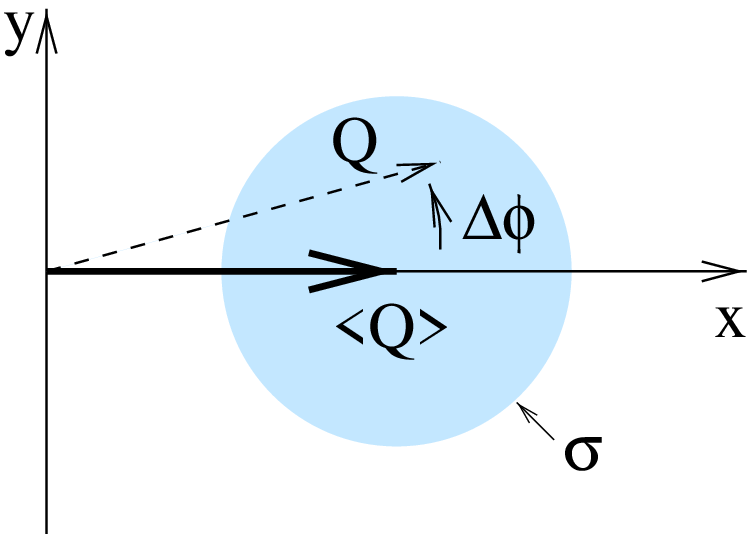}
\caption{Schematic picture of the distribution of ${\bf Q}$, 
given by Eq.(4). 
${\bf Q}$ fluctuates around its average value 
$\langle {\bf Q}\rangle =\bar Q\,{\bf e}_x$ 
with a standard deviation $\sigma$, so that a typical value of ${\bf Q}$ 
(dotted arrow) lies within the shaded circle with radius $\sigma$.}
\label{fig:fig4}
\end{minipage}
\hspace{\fill}
\begin{minipage}[t]{72mm}
\epsfbox{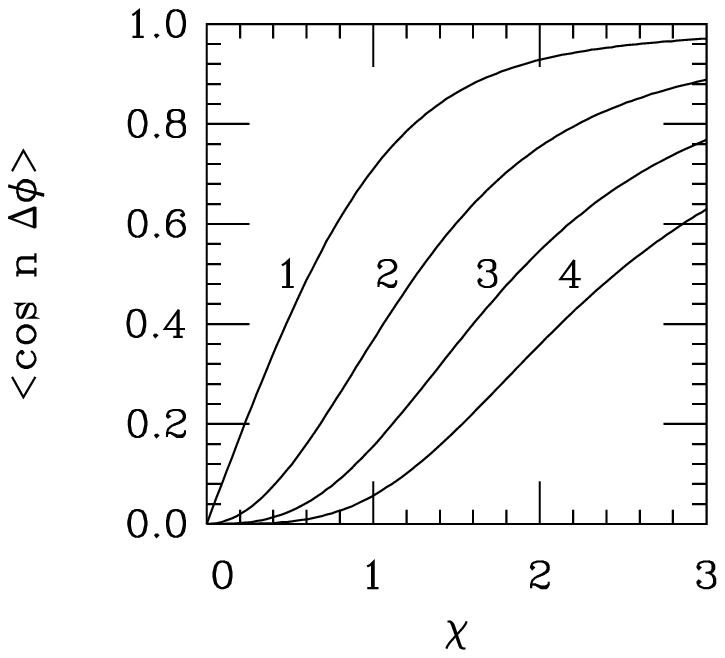}
\caption{Variation of 
$\langle\cos n\Delta\phi\rangle$ with the parameter $\chi$, calculated 
from Eq.(5). The curves are labeled by the value of $n$.}
\label{fig:fig5}
\end{minipage}
\end{figure}

The direction of impact parameter, estimated from Eq.(\ref{Q}), 
differs from the true one by an error $\Delta\phi$, 
due to finite multiplicity fluctuations (see Fig.\ref{fig:fig4}). 
The effect of this error on flow measurements must be corrected. 
I now explain a method \cite{Olli97} 
which allows to do this correction systematically. 

The measured azimuthal angle $\psi$ of an outgoing particle is related 
to the true azimuthal angle $\phi$ by $\psi=\phi-\Delta\phi$. 
Averaging over many events, assuming that $\phi$ and $\Delta\phi$ are 
independent (autocorrelations are avoided by removing the particle from 
the sum in Eq.(\ref{Q}) \cite{Dani85}), one obtains the following relation 
between measured and true Fourier coefficients \cite{Barr97a,Demo90,Olli97}:
\begin{equation}
\label{deconvolution}
\langle\cos n\psi \rangle  
=\langle \cos n\phi\rangle\langle \cos n\Delta\phi\rangle.
\end{equation}
From Eq.(\ref{deconvolution}), one can reconstruct the true distribution once 
the correction factor $\langle\cos n\Delta\phi\rangle$ is known. 

The important point is that, as we shall see shortly, the distribution 
of $\Delta\phi$ is a universal function of a single dimensionless parameter
$\chi$, which measures the accuracy of the reaction plane determination 
and can be determined directly from experiment. 

In order to calculate the distribution of $\Delta\phi$, we must 
parametrise the fluctuations of ${\bf Q}$. 
Since the number $N$ of particles entering the definition of the vector 
${\bf Q}$ in Eq.(\ref{Q}) is usually much larger than unity, the central
limit theorem can be applied. It shows that, 
for given magnitude and orientation of the true impact parameter, 
the fluctuations of $\mbox{\bf Q}$ around its average value 
$\langle{\bf Q}\rangle=\bar Q\, {\bf e}_x$ (which is the true 
direction of impact parameter) are gaussian. 
Then the two dimensional distribution of ${\bf Q}$ takes the form
\cite{Olli92,Olli93b,Volo96a,Olli95}
\begin{equation} 
\label{gaussian}
{dN\over Q dQ d\Delta\phi}={1\over\pi\sigma^2}
\exp\left(-{\left| {\bf Q}-\langle{\bf Q}\rangle\right|^2\over\sigma^2}\right)
={1\over\pi\sigma^2}
\exp\left(-{Q^2+\bar Q^2-2 Q\bar Q\cos\Delta\phi\over\sigma^2}
\right).
\end{equation}
Note that $\bar Q$ scales like $N$ while $\sigma$ scales like $\sqrt{N}$. 

Eq.(\ref{gaussian}) can be integrated over $Q$~\cite{Olli93b} to 
yield the distribution of $\Delta\phi$. 
As can be seen in Fig.\ref{fig:fig4}, the typical magnitude of 
$\Delta\phi$ is $\sigma/\bar Q$. It is therefore useful to introduce 
the dimensionless parameter $\chi\equiv\bar Q/\sigma$, which characterises
the accuracy of the reaction plane determination. 
The correction factors $\langle\cos n\Delta\phi\rangle$ entering 
Eq.(\ref{deconvolution}) depend on $\bar Q$ and $\sigma$ only through 
$\chi$. They are calculated by integration of 
Eq.(\ref{gaussian}) over $\Delta\phi$ and $Q$ \cite{Olli97}:
\begin{equation}
\label{pn}
\langle\cos n\Delta\phi\rangle=
{\sqrt{\pi}\over 2}\chi e^{-\chi^2/2}\left[
I_{n-1\over 2}\left( {\chi^2\over 2}\right) +
I_{n+1\over 2}\left( {\chi^2\over 2}\right) \right]
\end{equation}
where $I_k$ is the modified Bessel function of order $k$.
The variations of the first coefficients with $\chi$ is displayed 
in Fig.\ref{fig:fig5}. Since $\Delta\phi$ is of order $1/\chi$, 
$\langle\cos n\Delta\phi\rangle$ is close to unity for large $\chi$. 

The last step of our reconstruction procedure 
is to estimate the value of $\chi$, i.e. to measure 
the accuracy of the reaction plane determination.  
The most popular method to do so 
\cite{Croc97b,Just95,Bast97,Bril93,Vene93} 
is to divide each event randomly into two 
subevents containing half of the particles each, 
and to construct {\bf Q} independently for each subevent \cite{Dani85}. 
One thus obtains two vectors $\mbox{\bf Q}_I$ and $\mbox{\bf Q}_{II}$. 
The distributions of $\mbox{\bf Q}_I$ and $\mbox{\bf Q}_{II}$ 
are given by an equation similar to Eq.(\ref{gaussian}). 
However, since each subevent contains only $N/2$ particles, 
the corresponding average value and fluctuations must be scaled:
$\bar Q_I=\bar Q_{II}=\bar Q/2$, $\sigma_I=\sigma_{II}=\sigma/\sqrt{2}$, 
and therefore $\chi_I=\chi_{II}=\chi/\sqrt{2}$. 
The distribution of the relative angle between $\mbox{\bf Q}_I$ and 
$\mbox{\bf Q}_{II}$,
$\Delta\phi_R\equiv\left|\Delta\phi_I-\Delta\phi_{II}\right|$
can be calculated analytically \cite{Olli93b,Olli95}. 
The fraction of events for which $\Delta\phi_R>90^\circ$ 
is simply related to $\chi$: 
\begin{equation}\label{ratio}
{N(90^\circ<\Delta\phi_R<180^\circ)\over N(0^\circ<\Delta\phi_R<180^\circ)}
={\exp(-\chi^2/2)\over 2}.
\end{equation}
Once $\chi$ is known, the Fourier coefficients of the azimuthal 
distribution can be reconstructed from Eqs.(\ref{deconvolution}) and 
(\ref{pn}). 
Until now, only the two first components, $n=1$ (directed flow) and 
$n=2$ (elliptic flow) are measured. It would be interesting to extend 
these measurements to higher harmonics \cite{Volo96a}. 
Upper bounds on the next two Fourier coefficients have been given 
by E877 \cite{Barr97b,Barr94}. 

The transverse momentum method, Eq.(\ref{Q}), gives a reasonable 
estimate of the reaction plane if directed flow is large enough. 
However, directed flow is weak at SPS and will probably not be observable
at RHIC and LHC. For this reason, an alternative method to determine 
the reaction plane was proposed in \cite{Olli93b}, which makes 
use of elliptic flow, instead of directed flow. 
If in-plane elliptic flow is present, $\Phi_R$ is estimated according to 
the following equation: 
\begin{equation}
\label{Q2}
{\bf Q}_2 =\pmatrix{Q_2\,\cos 2\Phi_R\cr Q_2\,\sin 2\Phi_R}=
\sum_{k=1}^N {w_k\pmatrix{\cos 2\phi_k\cr\sin 2\phi_k}}. 
\end{equation}
In this case, the weight $w_k$ is positive everywhere, since the 
sign of elliptic flow does not depend on rapidity. 
This method is equivalent to a diagonalisation of the transverse 
sphericity tensor \cite{Olli92}, and was proposed independently 
in \cite{Wils92} for low energies where directed flow also vanishes.
Note that the last equation allows to determine $\Phi_R$ modulo $\pi$. 
The magnitude of elliptic flow at SPS is determined more accurately 
with this method than with the transverse momentum method \cite{Appe97}. 

Note that the above methods do not allow to determine the sign of the flow. 
Eq.(\ref{Q}) {\sl assumes\/} that directed flow is positive 
(if it was negative, $\Phi_R$ should be replaced by 
$\Phi_R+\pi$ in Eq.(\ref{Q})). Similarly, Eq.(\ref{Q2})
{\sl assumes\/} that elliptic flow is in-plane. 
In fact, the sign of directed flow was determined only 
at low energies, where it is negative, 
by coincident polarisation measurements \cite{Tsan86}. 
The sign of elliptic flow can be determined only by a simultaneous 
measurement of directed flow. 

One must be aware that the above description of flow analysis 
is oversimplified. 
In particular, Eqs.(\ref{Q}) and (\ref{Q2}) cannot be used in experiments 
where particles are not seen individually, but only through the 
energy deposited in a calorimeter, as in most AGS and SPS experiments. 
I refer the reader to the corresponding publications for more details 
\cite{Barr97a,Appe97,Barr97b,Barr94,Wien96}. 

\section{CONCLUSIONS AND OUTLOOK}

Directed and elliptic flow are present at all energies in nuclear collisions. 
Their measurement provides direct insight into final state interactions, 
which can only be deduced indirectly from other observables such as 
inclusive distributions. 
Therefore, flow is highly sensitive to medium properties. 
Detailed experimental analyses, carried out with large acceptance 
detectors, are now available at Bevalac and SIS energies. 
Since a given observable usually gets contributions from several 
effects, it is difficult in practice to extract accurate numbers for  
physical quantities such as compressibility, optical potentials, 
in-medium cross sections. 
Nevertheless, it has been recently emphasized that no model is 
presently able to explain all the data quantitatively \cite{Croc97b}, 
which shows that flow provides useful constraints on theory.

The increase of directed and elliptic flow with fragment mass indicates 
that they reflect the collective motion of the expanding system. 
On the other hand, their increase with the mass numbers of colliding 
nuclei indicates that there is no geometric scaling at SIS energies, 
which implies that thermalisation is only partial. 
The small magnitude of elliptic flow at SPS suggests that the same 
holds at SPS. 
Flow measurements in collisions of lighter nuclei 
could provide some insight into the highly debated issue of thermalisation 
at ultrarelativistic energies. 

A wealth of new results have been obtained since the last Quark Matter 
conference. Flow is now studied at AGS by E877 
\cite{Volo97b}, E895 \cite{Liu97} and E917 \cite{Ogil97}, 
and at SPS by NA45 \cite{Cere97}, NA49 \cite{Appe97}, NA52 \cite{Kaba97}
and WA98 \cite{Nish97}. 
Directed flow and in-plane elliptic flow are now clearly seen at 
AGS and SPS, and flow of identified particles is measured 
\cite{Liu97,Ogil97,Nish97,Appe97,Barr97b,Kaba97}. 
It would be interesting to extend these measurements to 
other particles such as multistrange baryons, dileptons, 
$J/\Psi$'s. This would further constrain enhancement/suppression 
mechanisms. 

While directed flow becomes smaller as the bombarding energy increases,
in-plane elliptic flow becomes larger. One can reasonably 
expect that the latter will be the only observable flow effect 
at RHIC and LHC. Since in-plane elliptic flow results from 
compression of the produced hadronic matter, its magnitude should 
give information on the pressure, i.e. of the equation of state. 

In order to obtain a consistent description of heavy 
ion collisions, flow analyses should eventually be correlated 
with other observables. 
For instance, the knowledge of the reaction plane should allow
to obtain a more detailed information on the space-time distribution 
from HBT, 2 particle correlations \cite{Volo96b,Wied98}. 
A unified description of radial, directed and elliptic flow 
could also shed some light on the issue of thermalisation \cite{Volo97a}.

\end{document}